\newcommand{\gev}{\, {\rm GeV}}

\newcommand{\beq}{\begin{equation}}
\newcommand{\eeq}{\end{equation}}
\newcommand{\bea}{\begin{eqnarray}}
\newcommand{\eea}{\end{eqnarray}}

\newcommand{\gsim}{\lower.7ex\hbox{$\;\stackrel{\textstyle>}{\sim}\;$}}
\newcommand{\lsim}{\lower.7ex\hbox{$\;\stackrel{\textstyle<}{\sim}\;$}}

\documentclass[aps,prev,twocolumn,preprintnumbers,%
               floatfix,nofootinbib]{revtex4}
\usepackage{graphicx}
\usepackage{mathrsfs}
\usepackage{amssymb}
\usepackage{verbatim}
\def\stacksymbols #1#2#3#4{\def\theguybelow{#2}
    \def\vp{\lower#3pt}
    \def\sp{\baselineskip0pt\lineskip#4pt}
    \mathrel{\mathpalette\intermediary#1}}

\def\intermediary#1#2{\vp\vbox{\sp
     \everycr={}\tabskip0pt
     \halign{$\mathsurround0pt#1\hfil##\hfil$\crcr#2\crcr
              \theguybelow\crcr}}}

\def\comment#1{}
\def\to{\rightarrow}

\def\u1x{U(1)_X}
\newcommand{\nc}{\newcommand}
\nc{\LL}{L} \nc{\vv}{\tilde{v}} \nc{\ccdot}{\!\cdot\!}
\nc{\gsm}{G_{SM}}
\nc{\vfive}{\mathbf{5}\oplus\mathbf{\overline{5}}}
\nc{\vten}{\mathbf{10}\oplus\mathbf{\overline{10}}}
\nc{\zhol}{Z^{\rm hol}}

\begin{document}

\preprint{~~~~CERN-PH-TH/2008-12; MCTP-08-5; UCI-TR-2008-3}

%\bigskip\bigskip

\title{Probing CP-violation at colliders through \\ interference effects in diboson production and decay}

\author{Jason Kumar$^{a}$}
\email{kumarj@uci.edu}
\author{Arvind Rajaraman$^{a}$}
\email{arajaram@uci.edu}
\author{James D. Wells$^{b,c}$}
\email{jwells@umich.edu} \vspace{0.2cm} \affiliation{
${}^a$ Physics Department, University of California, Irvine, CA 92697 \\
${}^b$ CERN, Theory Division (PH-TH), CH-1211 Geneva 23, Switzerland \\
${}^c$ Michigan Center for Theoretical Physics (MCTP),
 Ann Arbor, MI
48109}

\begin{abstract}

We define a CP-asymmetric observable that is sensitive to CP-violating interactions
in the gauge-boson sector. We illustrate the utility of this observable by studying how well
the LHC can measure the coefficient of
a particular dimension-six $WWZ$ operator.
We find that sensitivity at the $10^{-3}$ level is possible at the LHC with $100\, {\rm fb}^{-1}$ of
integrated luminosity, which would greatly exceed the sensitivity achieved at LEP, and
would rival or may even better the indirect sensitivities inferred from related operators
constrained by electric dipole moment experiments.

\bigskip\bigskip\bigskip

\end{abstract}

\maketitle

\setcounter{equation}{0}

%%%%%%%%%

%%%%%%%%%%%%%%%%%%%%%%%%%%%%%%%%%%%%%%%%%%%%%%%%%%%%%%%%%%%%%%%%%%%%%%

\begin{center}{\it Probing CP-violation}\end{center}

One of the most well motivated possibilities
for new physics is CP-violation.
Many new
experimental probes of CP-violation have been studied, both at
accelerators and at other experiments.
 There are several reasons for
this. Firstly, CP-violation has been observed in
kaon decays and there is great interest in determining all
possible theoretical sources of new physics which could contribute,
as well as possible new experimental signatures of
CP-violation.  Secondly, CP-violation is
required for baryogenesis.
The known source of CP-violation in the Standard Model (SM) -- the
CKM phase -- is not sufficient to generate the known
baryon asymmetry, and so some other source is needed.

In many models, large CP-violation can be induced in the gauge boson sector.
For instance, an exotic fermion coupled to the electroweak bosons can induce CP-violating
couplings. The large number of fermions that can arise in intersecting
brane models of string theory could thus be a source of large CP-violation
in the gauge-boson sector.
It is therefore of great interest to look for the effects of such new
physics (related triple gauge boson coupling signatures from string theory have
been discussed in~\cite{IBMTGC}).

In this paper, we discuss the possibility of probing CP-violation in the gauge-boson sector
 at colliders, and in particular, at the Large Hadron Collider (LHC).
We introduce observables that are directly
sensitive to CP-violation, and argue that they
can be utilized to probe CP-violating couplings at a wide variety of
accelerator
experiments, and for a large class of new physics models.
We apply this to a specific operator which contributes to the WWZ vertex, and show that collider
searches can improve current bounds on this operator by well over an order of magnitude.

%%%%%%%%%%%%%%%%%%%%%%%%%%%%%%%%%%%%
\begin{center}{\it CP violation in the WWZ triple gauge couplings}\end{center}

We begin by considering new physics that modifies the WWZ vertex.
The WWZ vertex can, up to general dimension six operators,
be parameterized in terms of the
effective Lagrangian~\cite{Gounaris:1996rz}
\bea
\label{operator}
i{\cal L}_{eff}= g_{WWZ}\Big[g_1^ZZ^\mu(W^{-}_{\mu\nu}
W^{+\nu}-W^{+}_{\mu\nu}
W^{-\nu})\nonumber\\+\kappa_Z
W^+_{\mu}W^-_{\nu}Z^{\mu\nu}
+{\lambda_Z\over m_W^2}Z^{\mu\nu}W^{+\rho}_{\nu}W^-_{\rho\mu}
\nonumber\\
+ig_5^Z\epsilon_{\mu\nu\rho\sigma}((\partial^\rho W^{-\mu})W^{+\nu}
-W^{-\mu}\partial^\rho W^{+\nu})Z^\sigma
\nonumber\\
+ig_4^ZW^-_{\mu}W^+_{\nu}(\partial^\mu Z^{\nu}+\partial^\nu Z^{\mu})
\nonumber\\
+{\tilde\kappa_Z \over 2}W^-_{\mu}W^+_{\nu}\epsilon^{\mu\nu\rho\sigma}
V_{\rho\sigma}
-{\tilde \lambda_Z \over 2M_W ^2}
W_{\rho \mu} ^- W^{+\mu}_{~~~\nu}
\epsilon^{\nu \rho \alpha \beta} Z_{\alpha \beta}\Big]
\eea
where $W_{\mu \nu} =
\partial_{\mu} W_{\nu} - \partial_{\nu} W_{\mu}$, $Z_{\mu \nu}
= \partial_{\mu} Z_{\nu} - \partial_{\nu} Z_{\mu}$.
In the SM, $g_1^Z=\kappa_Z=1$, and all the other terms are zero.

In this Lagrangian, $g_1^Z,\kappa_Z,\lambda_Z, g_5^Z $ are
CP-conserving, and the other terms are CP-violating. The CP conserving
operators have been studied in
great detail~\cite{CP conserving summary}, and the bounds on these parameters have been
analyzed (see e.g., LEP studies in~\cite{Alcaraz:2006mx}). The CP-violating operators have also
been studied at colliders~\cite{CP Violation Studies - Theory,CP Violation Studies - Experiment},
but the bounds are only at best $\sim 0.1$.
The DELPHI Collaboration~\cite{CP Violation Studies - Experiment}
used the process $e^+e^-\to W^+W^-\to l\nu q\bar q(l=e/\mu)$
to obtain the measurements
\bea
g_4^Z & = & -0.39^{+0.19}_{-0.20}, \\
\tilde \kappa_Z & = & -0.09^{+0.08}_{-0.05}, \\
\tilde\lambda_Z & = & -0.08\pm 0.07
\eea
LEP and Tevatron sensitivities to the related
coefficient $\tilde \lambda_{\gamma}$ are only at $\tilde
\lambda_{\gamma} \leq 0.3$~\cite{Abreu:1997wh,Abachi:1996hw}.

We will now consider the sensitivity to the LHC to these coefficients.
We consider a scattering process
with matrix element ${\cal M}_0+\delta{\cal M}$, where ${\cal M}_0$ is
the SM matrix element and $\delta {\cal M}$ is the
contribution arising from new physics.
The leading change in the cross-section due to new physics
is then the interference term
\beq \delta \sigma \propto \Re ({\cal M}_0 \delta {\cal M}^*).
\eeq

We now wish to look for CP-violating physics in the interference
effects.
We assume that the SM matrix element is CP-conserving;  this will
be the case in any process for which fewer than three generations
participate.  Even more generally, the only source of SM CP-violation
is the small contribution from the 
CKM phase, and we assume new physics to carry the
larger contribution. This assumption is especially warranted if we
envision the new CP-violation as accounting for the baryon asymmetry.

To look for the effects of new physics, we
note that  a term in the cross-section
proportional to $\epsilon_{\mu \nu \rho \sigma}$
is always a signal of CP-violating
physics. One way to see this is that such a term is odd under
naive time reversal
(the flip $t\rightarrow -t$). This suggests that it will probe a CP-violating
term. Indeed explicit computations using the effective
Lagrangian (\ref{operator}) show that all terms
proportional to the epsilon tensor in the interference term are proportional to
CP-violating coefficients.
Note that $g_5^Z$ is the
coefficient of a parity violating, CP-conserving operator which also
is proportional to the $\epsilon$ tensor.  But because this term
comes with an imaginary coefficient, it
will cancel out of the interference
cross-section.

We will therefore focus on terms in the
cross-section proportional to $ \delta \sigma \propto
\epsilon_{\mu \nu \rho \sigma} $.
Experimental signals of these terms  can be used to probe the
couplings $g_4^Z,\tilde\kappa_Z, \tilde\lambda_Z$.
In this note, we shall  discuss the
experimental sensitivities on  $\tilde\lambda_Z$, leaving the more exhaustive analysis
for future work.

%%%%%%%%%%%%%%%%%%%%%%%%%%%%%%%%
\begin{center}{\it Signals of CP-violation}\end{center}

One can write the first-order shift in the differential cross-section for the
process $q \bar q \rightarrow W^* \rightarrow WZ \rightarrow
l \nu Z$ as
\bea
 d \sigma &=&{1\over 12} {1\over
2 E_q 2E_{\bar q} |v_q - v_{\bar q}|}
\left(\prod_{f=l,\nu,Z}
{d^3 p_f \over (2\pi)^3} {1\over
2E_f}\right)
\nonumber\\
&\,& \times
(2\pi)^4 \delta^{4}(P+\sum p)
\times  \Re(2{\cal M}_0 \delta {\cal M}^*).
\eea
The SM matrix element ${\cal M}_0$ is given
by $W$,$Z$ production via $t$- and $u$-channel exchange
of a quark, and by $s$-channel production of an off-shell
$W^*$ boson decaying to $W$, $Z$ via the
SM WWZ vertex
\bea
\Gamma_{\mu
\nu \rho} = {i {e \cot\theta_W}}(k_{1 \mu} g_{\nu \rho} -k_{1\rho}
g_{\nu \mu} -k_{2\mu} g_{\nu \rho} \nonumber\\
+k_{2\nu} g_{\rho \mu}
+k_{Z\rho} g_{\mu \nu} -k_{Z\nu} g_{\mu \rho})
\eea
Here
$k_{1,2}$ are  the momenta of the $W$'s and $k_Z$ is the momentum of
the $Z$.

If $\tilde\lambda_Z$ is nonzero,
the WWZ vertex is shifted by a term of the form
\bea
\label{vertex}
\delta\Gamma_{\mu
\nu \rho} &=& e \cot \theta_W {\imath \tilde \lambda_Z \over M_W
^2} (k_{2\nu} \epsilon_{\mu \rho \sigma \tau} k_2 ^{\sigma} k_1
^{\tau} +k_{1\rho} \epsilon_{\mu \nu \sigma \tau} k_2 ^{\sigma}
k_1 ^{\tau}
\nonumber\\
&\,&-k_1 \cdot k_2 \epsilon_{\mu \nu \sigma \tau} k_2 ^{\sigma}
k_1 ^{\tau}) \eea
This vertex will lead to a potentially observable correction
to the cross-section for WZ production at the LHC.

The immediate difficulty is that a spin-averaged
$2 \rightarrow 2$ scattering
process cannot yield a term in the cross-section proportional to
the epsilon tensor.
This is because there are only 3 independent momenta in a $2
\rightarrow 2$ process, while the $\epsilon$ contribution will be
non-zero only if contracted into 4 independent momenta.   For
example, one cannot detect an asymmetry in the
spin-averaged process $q \bar q \rightarrow WZ$.

To obtain an asymmetry, one must keep track of the polarization of
the outgoing gauge bosons. There is a vast literature on measuring
$W$- and $Z$-polarizations, via asymmetries in their decays to leptons
or jets.
A complete analysis using these polarizations is left for future
work. For this analysis, we shall instead focus on a particular
decay channel $W\rightarrow l\nu, Z\rightarrow ll$ which has a
clean trilepton signal.
This will enable us to use the background analysis of \cite{DobbsLefebvre}.

Specifically, we
denote by $p_q$ and $p_{\bar q}$ the momenta of the
incoming quark and antiquark respectively, and by $p_l$ and
$p_{\nu}$
the momenta of the lepton and neutrino arising from the decay of the
outgoing $W$. We treat the $Z$ as an outgoing
particle with momentum $k_Z$, since it can be
reconstructed easily using the $Z\to l^+l^-$ decay product
leptons.
Then we will have new terms in the cross section
proportional to
\beq \epsilon_{\mu \nu \rho \sigma}
(p_q + p_{\bar q} )^{\mu} (p_q - p_{\bar q} )^{\nu}
p_l ^{\rho} k_Z ^{\sigma}
\eeq
As explained above, such a term is a
direct probe of CP-violation.

For the form of $d\sigma$ given above, the
integrated change in the cross-section will vanish. To obtain a
nonzero result, we must weight
the events by an asymmetric
observable which is itself parity-asymmetric,
for instance, %we can weight events by
the sign of a
triple-product.
We further observe that $p_q$ and $p_{\bar q}$ have
non-zero components only along the time and beam axes.  This
implies that the outgoing lepton and $Z$ contraction into the
$\epsilon$ is proportional to $k_Z ^T \times p_l ^T$. Hence, for our
asymmetric observable, we should weight events by the sign of the
cross-product $p_q \cdot (k_Z \times p_l)$.

But we cannot measure the momentum of the quark, and there is
a 4-fold ambiguity in its reconstruction.
We will instead use the momentum of the $Z$ along
the beam axis as a proxy for the quark  momentum. Since the quark
typically has a larger momentum fraction than the anti-quark, the
$Z$-boson will typically move in the same direction along the
beam axis as the quark. Through numerical simulations we find that this
correlation is $\gsim 70\%$, so the CP-asymmetry
will not be degraded significantly
by choosing the  $Z$ momentum as the proxy for the quark momentum.

We will therefore weight events by
\begin{equation}
\Xi^z_{\pm}(k_Z,p_l) \equiv
sgn(k_Z ^z) sgn(p_l \times k_Z)^z
\end{equation}
as a substitute for the more direct, but unmeasurable full triple
product.  Although this substitution
is imperfect, it should provide for a non-vanishing weighted
cross-section and a striking test of CP-violation if it is present.
The resulting  asymmetric observable is then obtained
by integrating the sign-weighted differential cross-section
\begin{equation}
\label{observable}
\Delta\sigma =\int  d\sigma(pp\rightarrow W^* \rightarrow WZ)
\, \Xi^z_{\pm}(k_Z,p_l)
\end{equation}
Experimentally, this observable is measured by counting
trilepton events weighted by a sign determined from the observed
momenta.

\begin{center}{\it Event rates}\end{center}

Considerable effort has been expended in determining the ability of
the LHC to probe corrections to the $WWZ$ vertex, particularly
through the $pp \rightarrow WZ \rightarrow lll\nu$ channel.
We can
therefore make use of the cuts and backgrounds determined for
previous WWZ analyses.

We will here follow the analysis presented in~\cite{DobbsLefebvre}.
The following cuts were used in this analysis:
\begin{itemize}
\item{Three isolated electrons or muons with $|\eta| <2.5$ and
$|P_T| >25\gev$.}
 \item{Two leptons are of like flavor and
opposite sign, and reconstruct to an on-shell $Z$ within $10\gev$.}
\item{Missing $P_T > 25\gev$}
\item{No other charged leptons
with $|\eta|<2.5$, $|P_T|>25\gev$}
\item{There exists a solution
for neutrino momentum that reconstructs to an on-shell $W$}
\end{itemize}
Subject to these cuts, the number of
events with $30\, {\rm fb}^{-1}$ of integrated
luminosity was found to be $\sim 2500$, including both tree-level
$WWZ$ processes and other SM contributions~\cite{DobbsLefebvre}.

These events will be distributed symmetrically. We
therefore expect to have $\sim 1250$ events with one particular sign of the
$\Xi^z_\pm(k_Z,p_l)$, and $\sim 1250$ events with the opposite sign. The net
expected value of the observable $\Delta\sigma$ is thus zero. However, due to
the statistical uncertainties, the observable will have a variance
of $\sqrt{2500}\sim 50$. To have a signal-to-background ratio of
5, we  need $\sim 250$ asymmetric events with $30\, {\rm fb}^{-1}$, and by
extrapolation, $\sim 460$ asymmetric events with $100\, {\rm fb}^{-1}$ of data.

Note that the number of asymmetric events required is still only
$\sim 10\%$ of the number of tree-level events.  This is
consistent with a small linear asymmetric correction, where the
quadratic piece can be ignored when computing the statistical significance
of the $\Delta\sigma$ asymmetry signal.

%%%%%%%%%%%%%%%%%%%%%%%%%%%%%%%%%%%%%%
\begin{figure}[t]
\centering
\includegraphics[width=8cm]{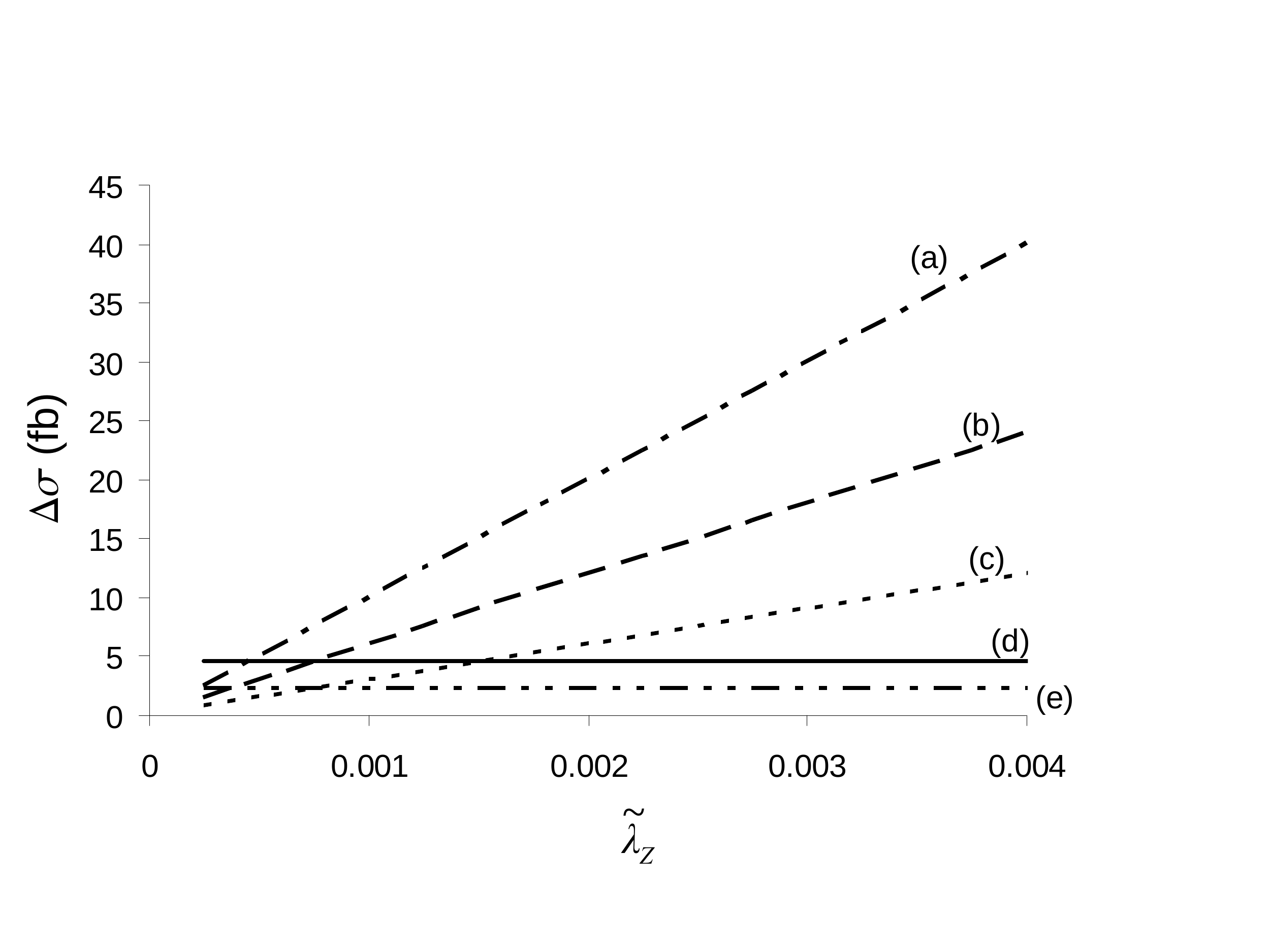}
\caption{Plot of $\Delta\sigma$ asymmetry cross-section as a function of $\tilde\lambda_Z$.
Lines (a), (b) and (c) correspond respectively to $\Delta \sigma$ in the
cases where no kinematics cuts are imposed, kinematic cuts on the
$Z$ decay products are imposed, and the full kinematic cuts are
imposed.  Lines (d) and (e) correspond to the
required $\Delta\sigma$ for $5\sigma$ and $95\%$ confidence
reach, respectively.}\label{crosssec}
\end{figure}
%%%%%%%%%%%%%%%%%%%%%%%%%%%%%%%%%%%%%%%%%%%

\begin{center}{\it Results}\end{center}

 We can now calculate the reach of the LHC for the
vertex (\ref{vertex}).
We compute the linear interference term in the
$pp \rightarrow W^* \rightarrow WZ$ cross-section by
computing the Feynman diagrams associated with $q\bar q'\to WZ$. There are
four such diagrams, three of which are SM diagrams ($s$-channel $W^*$ exchange,
and $t$ and $u$-channel quark exchange diagrams), and one is the CP-violating
interaction diagram ($s$-channel $W^*$ exchange with CP-violating $WWZ$ interaction).
We then generate a large number of events using PYTHIA 6.401~\cite{Pythia}, modified to
include the CP-violating interaction and the weighted signs $\Xi^z_\pm(k_Z,p_l)$.
We calculate the cross-section for the asymmetric observable at the LHC to be
\beq
\Delta\sigma \simeq \tilde \lambda_Z \times  (3 \times 10^3\, {\rm fb}).
\eeq

As shown above, we need $\sim 460$   asymmetric events for a $5\sigma$ detection of this
operator with  $100\, {\rm fb}^{-1}$ of data.
We conclude that LHC should be sensitive to the $\tilde \lambda_Z$
operator coefficient at the level of
\beq
\tilde \lambda_Z \lsim 0.002
\eeq
with  $100\, {\rm fb}^{-1}$ of data.
This is almost two orders of magnitude better than the results of the
LEP2 experimental measurements.

The level of sensitivity is similar to the sensitivity that EDM experiments
have to $\tilde \lambda_\gamma$ and
$\tilde\kappa_\gamma$~\cite{Marciano:1986eh,Boudjema:1990cf,NovalesSanchez:2007th},
the coefficients of
related  CP-violating operators.  The sensitivity limits there are
approximately $|\tilde \kappa_{\gamma}|< 5.2\times 10^{-5}$
and $|\tilde \lambda_{\gamma}|< 0.019$~\cite{NovalesSanchez:2007th}.
Although $\tilde \lambda_Z$ is the coefficient of
 a different operator, it is often thought that
limits on any CP-violating operator apply to the rest of the operators
since they are presumably 
related by the underlying theory.  We have no strong opinion on this connection, but merely
note here that under this philosophy the LHC sensitivity rivals or may better that of EDMs.

\begin{center}{\it Additional Applications}\end{center}

In this letter we have  illustrated the general features of an
interference analysis which is very sensitive to
CP-violating physics.
The interference analysis we presented can be applied to a wide variety of
processes at different experiments.  For example, CP-violating
corrections to the WWZ vertex can also be studied at the
Tevatron~\cite{Tevatron studies},
through the process $p \bar p \rightarrow W^* \rightarrow WZ
\rightarrow lll\nu$. Although
the number of events in the sample is  currently low -- approximately 13 candidate events in
$1\,{\rm fb}^{-1}$ at D0~\cite{Tevatron experiment} --
some useful bounds may be obtainable if the luminosity
increases substantially and the CDF and D0 experiments are combined.

Similarly, one can probe CP-violation in the $WWZ$ vertex at
linear colliders~\cite{ILC studies} via the process $e^+ e^- \rightarrow Z^*
\rightarrow W^+ W^- \rightarrow l\nu l \nu$.  Note that the natural
channel for observing this effect operates when running well above
the $Z$-boson pole. As such, this type of analysis could provide a
very sharp tool at the ILC.

One can furthermore study a variety of similar
CP-violating operators at the LHC, such as $\tilde \lambda_{\gamma}$.
Due to the comparable efficiency in detecting the $\gamma$ as opposed to the $Z$,
one expects that the sensitivity to this operator at hadron colliders is
similar to the sensitivity to $\tilde \lambda_Z$.
However, one would have to consider the background in
detail in order to assess the detection possibilities.

Lastly, one can certainly probe CP-violation beyond the $WWZ$
and $WW\gamma$
vertices using this type of interference effect. For example,
CP-violation in the Higgs sector can manifest itself in $H^*\rightarrow ZZ$
decays after applying  a similar
analysis.

These channels are currently under study, and we hope to report on
them soon.

\noindent
{\it Acknowledgments:}
 We gratefully thank B. Dutta, S.
Bar-Shalom and C.-P. Yuan for useful discussions. This work is
supported in part by the Department of Energy.
The work of AR and JK is supported in part by
NSF Grants No.~PHY--0354993 and PHY--0653656. JK is grateful to Texas
A\&M University, where part of the work was done, for its hospitality.

\end{document}